# Using a double-frequency RF system to facilitate on-axis beam accumulation in a storage ring

B. C. Jiang [1], Z. T. Zhao[1,2*], S. Q. Tian[1], M. Z. Zhang[1], Q. L. Zhang[1]

1. Shanghai Institute of Applied Physics, Chinese Academy of Sciences, Shanghai 201800, China
2. ShanghaiTech University, Shanghai 201210, China

## ABSTRACT

An on-axis injection scheme using a double-frequency RF system in a storage ring with small dynamic aperture is proposed. By altering RF voltages, empty RF buckets can be created which will be used for on-axis injection. After bunches are injected, a reverse RF voltage altering process is performed and the injected bunches can be longitudinally dumped to the main RF buckets. The scheme allows reaping the advantages of the on-axis injection while still performing accumulation.

## I.　INTRODUCTION

A storage ring using "multi-bend achromat" (MBA) optics to reduce the natural emittance by one or two orders of magnitude down to current storage rings has been proposed since 1995 [1]. After decades of efforts, this will become reality at MAX IV lab [2]. If the ring circumference is in the 1~2 kilometers range, the emittance of the electron beam can reach the hard X-ray diffraction-limited region, an area that is of great interests to the synchrotron light community. Such a storage ring could provide synchrotron radiations of high brightness at a high repetition rate while holding a large percentage of spatially coherent fluxes [3, 4].

MBA optics employs much more numbers of bending magnets to decrease the bending angle $\theta$ of a single bending magnet to reduce the natural emittance which scales as the third power of $\theta$. For MBA optics, stronger quadrupoles are required to suppress the small dispersions created by the bending magnets, leading inevitably to the large negative native chromaticities in both transverse planes. In order to combat the head-tail instability and avoid large tune shifts of off-momentum particles, the chromaticities are usually corrected to slightly above zero. Thus strong chromatic sextuples are needed. These nonlinear elements will significantly reduce the dynamic apertures, present a great challenge for MBA lattice design [5, 6, 7]. Many of the MBA lattices which have been designed to date only provide horizontal dynamic apertures of around 2 mm [8, 9, 10], leading to severe difficulties for injection[11,12].

One of the on-axis injection schemes, "swap-out injection", has been proposed [13], which uses a fast dipole kicker to inject fresh high charge beam onto the closed orbit while the stored beam is extracted.

There are two ways to accomplish swap-out injection. One way is using an accumulation ring (AR) to exchange bunches with the main ring [14, 15]. The AR is filled by an injector such as a linac or a booster. The other way is to directly use a linac or a booster as an injector, with the swap-out beams from the main ring discarded. Swap-out scheme provides a baseline to inject the beam into the storage ring holding a rather small dynamic aperture without major technical challenges. However swap-out with an AR needs extra budgets to build and operate the AR. While for swap-out without an AR, it requires an injector capable of producing high and stable charged single bunch, also the scheme will get unfavorable radiation dose from the dumped beams. Recently, a novel longitudinal on-axis injection scheme has been proposed in [16], in which a low frequency RF system is used, the bunch is on-axis injected at the RF phase far away from the synchrotron phase with energy slightly higher than the stored beam. Then the injected bunch longitudinal damps to the synchrotron phase. For a storage ring with large energy acceptance, this is a compact on-axis injection scheme without significant changes of the hardwares.

* zhaozt@sinap.ac.cn

In this paper, a new on-axis injection scheme for the storage ring with small dynamic aperture is proposed, which using a double-frequency RF system and eliminating the need to swap-out stored beams. In the scheme, by altering two RF voltages, empty RF buckets will be created which can be taken for on-axis injection. After bunches are injected, the voltage altering process will be reversed and the injected bunches can be longitudinally transferred to the main RF buckets (where the stored bunches are located). The energy oscillation of the injected bunches in this process can be controlled to less than 1%. This injection scheme will be more tolerate to the energy jitter of the injected beam and the beam dynamics will be more controllable comparing to the scheme proposed in [16].

A closing variation injection process can be traced back to 1970s at Novosibirsk VEPP-3 positron storage ring [17], which holds 2 RF systems, one is at 1st harmonic and the other at 19th harmonic. 19 bunches are injected initially to collect enough positrons and then eventually stacked into one bucket as 1st harmonic RF system powered on. The difference for the scheme proposed in this paper is that it is shaped for top up injection for the light sources instead of collecting particles purpose and the longitudinal transient beam dynamic is studied which may be a concern for a light source. Also the second harmonic RF system can be used for bunch lengthening purpose as will be discussed.

In Section II we will explain how the empty buckets are created. In Section III, the longitudinal oscillations of the injected bunch and the disturbance to users will be investigated. Bunch length evolution, intra-beam scattering effects and the momentum acceptance variation during the injection process will be discussed in Section IV.

## II. TWIN RF BUCKETS PRODUCTION

Assume that the storage ring has two RF systems, with the main RF system operating at 250MHz and the other one at the 2nd harmonic frequency 500MHz. The time dependence of the RF voltages are given by

$$V_m = \widehat{V_m}(st)cos(\omega_m t) \qquad (1)$$

$$V_h = \widehat{V_h}(st)cos(2(\omega_m t - \varphi_s) - \frac{\pi}{2} + \Delta\theta). \qquad (2)$$

$$\varphi_s = arccos(\frac{U_0}{\widehat{V_{m0}}}), \qquad (3)$$

Where $V_m$, $V_h$ are the voltages of the main and the 2nd harmonic RF system respectively. The amplitudes of the RF voltages $\widehat{V_m}(st)$, $\widehat{V_h}(st)$ are modulated in a preset pattern. $\varphi_s$ is the synchrotron phase of the main RF system. $st$ denotes the modulating step, the voltage change between steps can be a linear ramp. Actually it is a continuous voltage modulation process, the steps are marked out for discussion convenience. $U_0$ is the radiation energy loss per turn, and $\omega_m$ is the angular frequency of the main RF system, $\widehat{V_{m0}}$ is the maximum RF voltage of the main RF system, $\Delta\theta$ is the synchrotron phase deviation.

Table 1. $V_m$ and $V_h$ variations as a function of step number.

| step | 1 | 2 | 3 | 4 | 5 | 6 | 7 | 8 | 9 |
|---|---|---|---|---|---|---|---|---|---|
| $\widehat{V_h}(st)/\widehat{V_{m0}}$ | 0.3 | 0.4 | 0.6 | 0.8 | 1 | 1 | 1 | 1 | 1 |
| $\widehat{V_m}(st)/\widehat{V_{m0}}$ | 0.8 | 0.8 | 0.8 | 0.8 | 0.8 | 0.6 | 0.4 | 0.2 | 0.1 |

If RF voltages are modulated by steps as shown in table 1, the total RF voltage evolution from step 1 to step 5 will be shown in Fig. 1. The stored bunch marked by the solid ellipses will be pushed backward and a second empty RF bucket marked by the hollow ellipses will be created for on-axis injection.

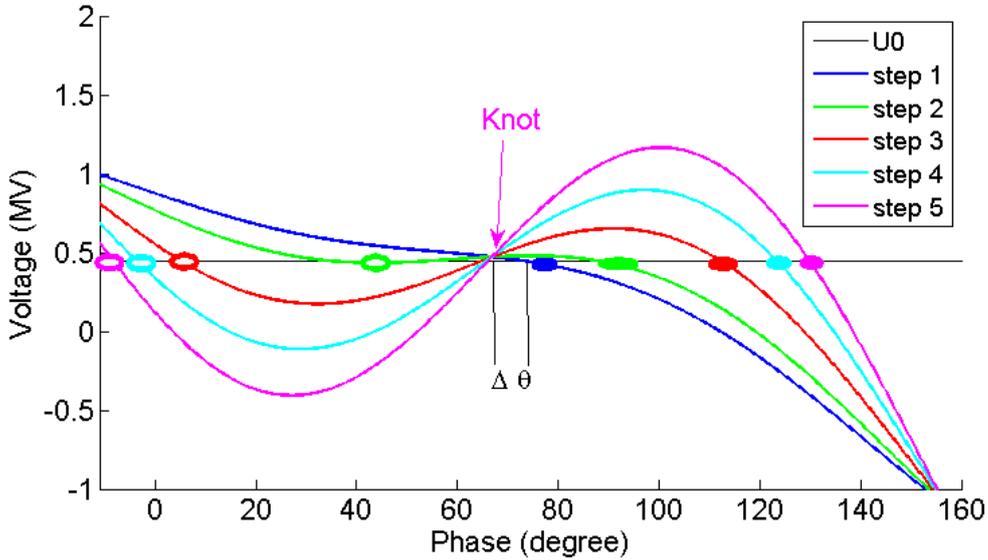

Fig. 1. Twin RF buckets creation by a double-frequency RF system with synchrotron phase deviation.

The departure of the 'knot' marked in Fig. 1 from the synchrotron phase is determined by $\Delta\theta$. A positive and sufficient large $\Delta\theta$ will ensure the stored bunch to be pushed backward entirely, prevent it being partitioned to the created bucket as will be discussed in detail in Section III. The separation of the twin buckets at step 5 as shown in Fig. 1 is about 1.5 ns. RF voltage modulation from step 6 to step 9 is to increase the twin buckets separation. At step 9 the separation is around 2ns. Fig. 2 shows the RF waveform of the whole process. At step 9 an on-axis injection can be performed with a 2ns pulse duration TEM-mode kicker [18] aimed at the empty bucket without swap-out stored beams.

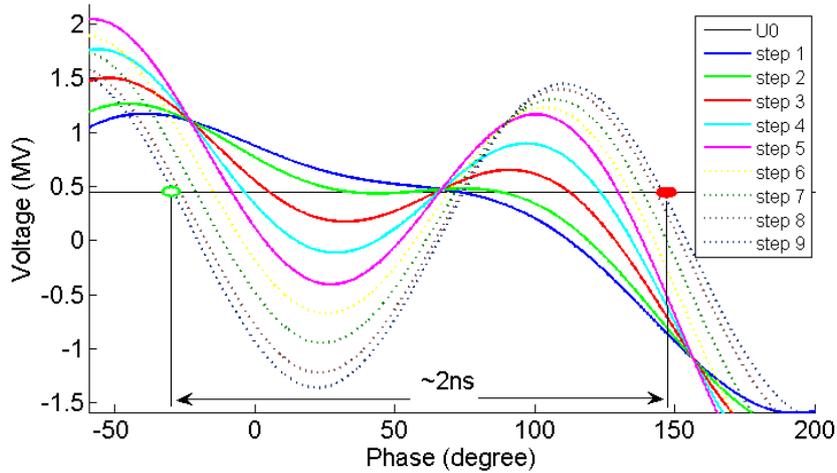

Fig. 2. Twin buckets separation.

For table 1, there are two points need to be explained. The first one is the RF voltage during the modulation process should avoid zero, because it is impossible to detuning the RF cavity to such a status. The second one is the starting step 1 with $\widehat{V_h}(1)/\widehat{V_{m0}} = 0.3$, $\widehat{V_m}(1)/\widehat{V_{m0}} = 0.8$ instead of $\widehat{V_h}(1)/\widehat{V_{m0}} = 0.1$, $\widehat{V_m}(1)/\widehat{V_{m0}} = 1$ is for bunch lengthening purpose, which will reduce the intra-beam scattering effect as will be shown in Section IV.

## III. DUMPING OF THE INJECTED BUNCHES TO THE MAIN RF BUCKETES

Here we skip discussions on the transverse beam dynamics of on-axis injection using a fast dipole kicker, which can be found elsewhere [15, 18]. We will focus on the longitudinal beam dynamics of the injection process.

After bunches are injected and captured by the storage ring, a reverse RF voltage altering process will be executed. Assuming that the voltage altering period $T_{RF}$ is on the order of seconds (much longer than synchrotron period $\tau_s$) no sudden phase jump will occur, and the twin RF buckets will approach each other smoothly. The key point is from step 2 to step 1 as indicated in Fig. 1. At this moment the injected bunch will be no longer stable in the longitudinal phase space. It will experience synchrotron oscillations and be eventually merged into the main RF buckets. The oscillation amplitude should be minimized to lessen the impact on the users and should also be within the energy acceptance of the storage ring.

The oscillation amplitude is relevant to the synchrotron phase deviation $\Delta\theta$. The smaller the value of $\Delta\theta$ the smaller of the energy oscillation amplitude will be. However $\Delta\theta$ should be sufficiently large so as not to split the stored bunch nor slip the store bunch forward to the created bucket. Taking the SSRF-U [19, 20] parameters as an example, the stored bunches occupy 6 degrees in phase (1.96 times the RMS bunch length, 95% of the total beam charge), plus 5 degree of synchrotron phase shift caused by double-frequency RF system at step 1 ( $\widehat{V_h}(1)/\widehat{V_{m0}} = 0.3$, $\widehat{V_m}(1)/\widehat{V_{m0}} = 0.8$), and plus 0.5 degree phase jitter for both RF systems as safety margins, implies that $\Delta\theta$ should exceed 12 degrees (degree calculation based on 500MHz).

Longitudinal tracking during the dumping process between step2 to step 1 can be performed by using Accelerator Toolbox in Matlab environment with a modified script of the RF voltage received by the particles.

$$V^n = \widehat{V_m}(st)cos(\varphi^n) + \widehat{V_h}(st)cos(2(\varphi^n - \varphi_s) - \frac{\pi}{2} + \Delta\theta) \quad (4)$$

where $n$ is the revolution turns, with $\Delta\theta = 12$ degree.

In the tracking, SSRF-U lattice is used whose main parameters are listed in table 2 and the optics is as shown in Fig. 3. The RF voltage is assumed to be steady state at every voltage step because $T_{RF}>>\tau_s$ and $T_{RF}>>\tau_f$. Where $\tau_f$ is filling time of the cavity $\tau_f=Q_L/(2\pi \cdot f_{rf})$.

2000 macro particles are generated randomly with a normal distribution in longitudinal phase space with energy spread 0.08%. At step 2 about 5% of the particles slip to the main RF bucket as plotted in Fig 4-a by cyan dots. Which indicates the injected bunch is separated into two parts, marks a beginning of the dumping process. A status between step 2 and step 1 with $\widehat{V_h}(st)/\widehat{V_{m0}} = 0.39$, $\widehat{V_m}(st)/\widehat{V_{m0}} = 0.8$ has also been tracked, it shows that at this moment all the particles slipped to the main RF bucket with energy oscillations less than 0.5% as shown in Fig 4-b which means the dumping process is finished.

Table 2. Main parameters of SSRF-U

| Lattice | 7BA |
|---|---|
| E$_0$ (GeV) | 3.0 |
| C (m) | 432 |
| ν$_{x/y}$ | 47.19/12.13 |
| V$_{RF}$ (MV) | 1.5 |
| Emittance (pm) | 205pm |
| Transverse emittance ratio | 10% |
| Damping time (x,y,z) (ms) | 11.1,19.7,16.2 |
| Bunch charge | 0.4mA/bunch (3.6e9 electrons) |
| Momentum compaction factor | 2.2e-4 |
| Energy spread | 0.08% |
| U$_0$ (MeV) | 0.438 |
| Bunch length (mm) | 3.5 （@500MHz） |
| Main RF frequency (MHz) | 250 |
| 2$^{nd}$ RF frequency (MHz) | 500 |
| R/Q | 90 Ω (for both RF cavities) |
| Q$_L$ | 1.3e5 (design)\ 1.65e5 (250MHz \ 500MHz) |

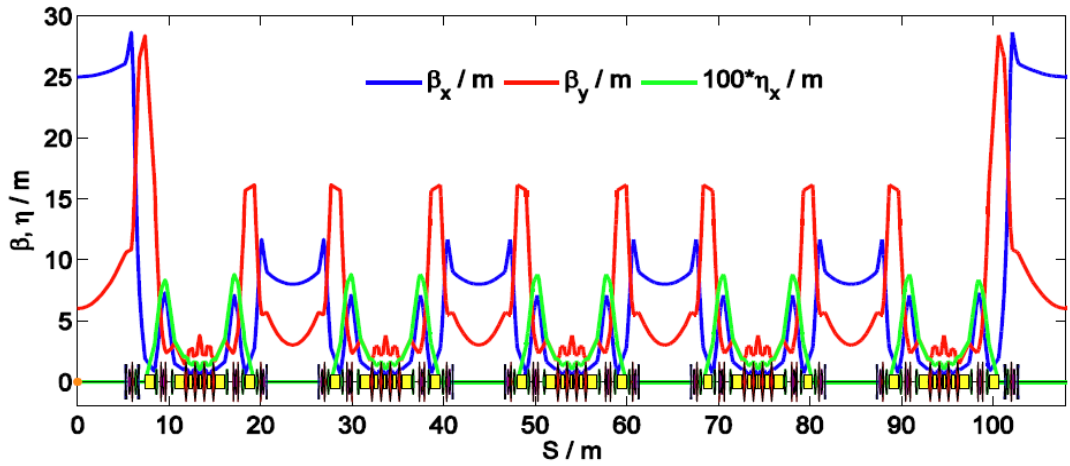

Fig. 3 Lattice and twiss parameters of a quarter of the SSRF-U.

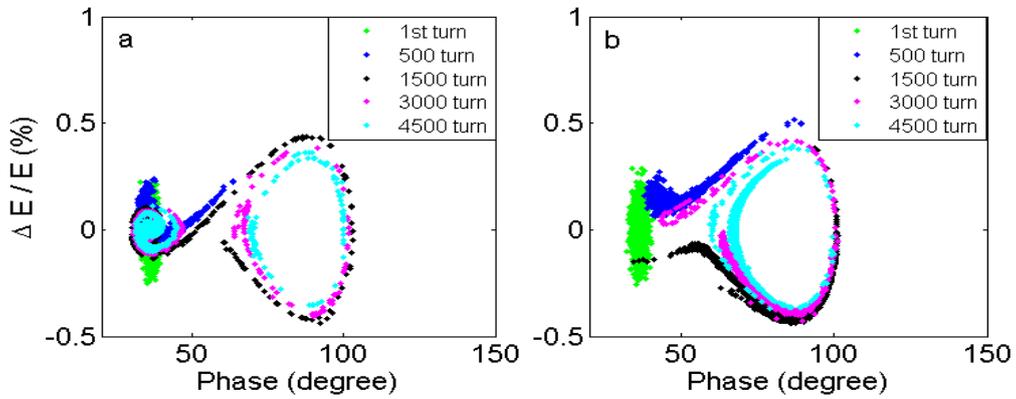

Fig 4. Multi particles tracking of dumping process.

The transverse oscillation of the injected bunch viewed at the straight section is negligible. Even at the maximum dispersion point in the arc, the oscillation amplitude of the injected bunch centroid is about 120 μm as shown in Fig 5.

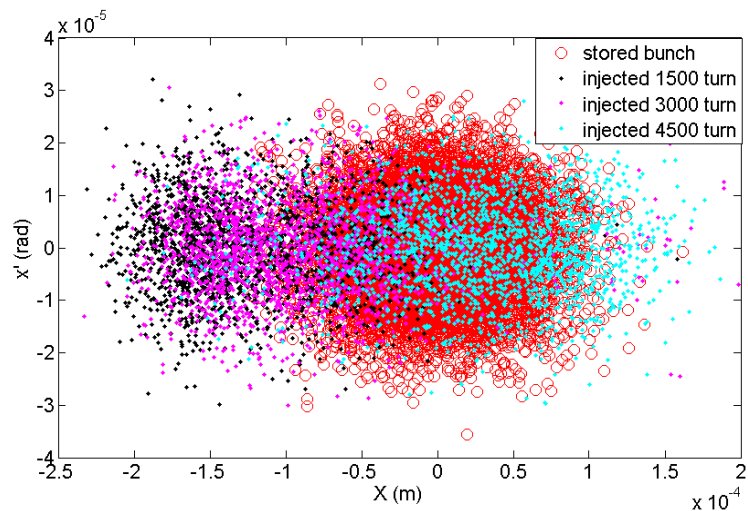

Fig. 5. Tracking of the transverse motion of injected bunch (snapshot at the maximum

dispersion point).

## IV. CALCULATION OF BUNCH LENGTH AND RF ACCEPTANCE EVOLUTION

In a storage ring with MBA lattice, the intra-beam scattering effect (IBS) will strongly influence the equilibrium emittance [2, 21]. IBS is highly sensitive to the bunch charge intensity. As RF voltages are modulated in the process, the stored bunch length will change during the period. As a result, the bunch charge density and the IBS growth rate will be modified. To evaluate the effects of IBS, the bunch length evolution is calculated for a combined voltage of $\widehat{V_m}(st)$ and $\widehat{V_h}(st)$ as listed in table 1. The stored bunch length evolution during the RF voltage altering process is shown in Fig. 6. It can be found out that with a proper setting as step 1 in table 1, the 2$^{nd}$ harmonic cavity can also act as a bunch lengthening system with a lengthening factor more than 3.

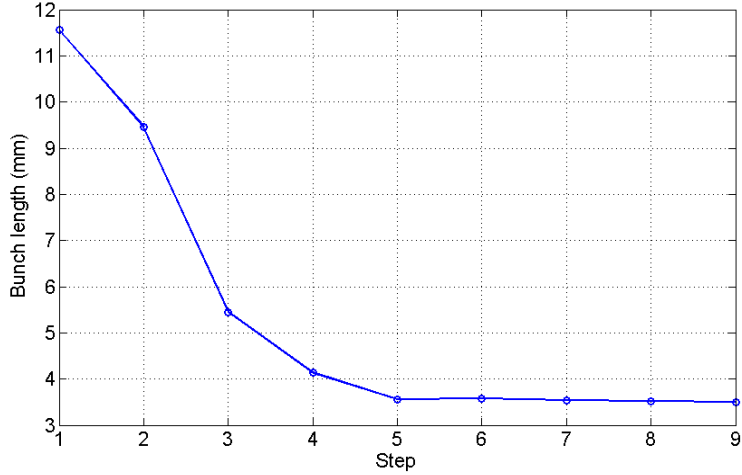

Fig. 6. Bunch length evolution of stored bunch.

With a given bunch length as shown in Fig. 6, the IBS is evaluated by the code MAD-X [22] which is based on Bjorken-Mtingwa formalism. Fig. 7 shows the equilibrium horizontal emittance counting IBS effect with transverse emittance ratio 10% ($\varepsilon_x$=185pm rad, $\varepsilon_x$=18.5pm rad). During the injection process the emittance fluctuates about 30% which is caused by the IBS effects. While for a 100% transverse emittance ratio ($\varepsilon_x$=103pm rad, $\varepsilon_x$=103pm rad), the emittance fluctuation will be 15%. The fluctuation will surely affect user experiments. A conservative way to mitigate this effect is sending a gating signal to the experimental hall to screen the data during injection.

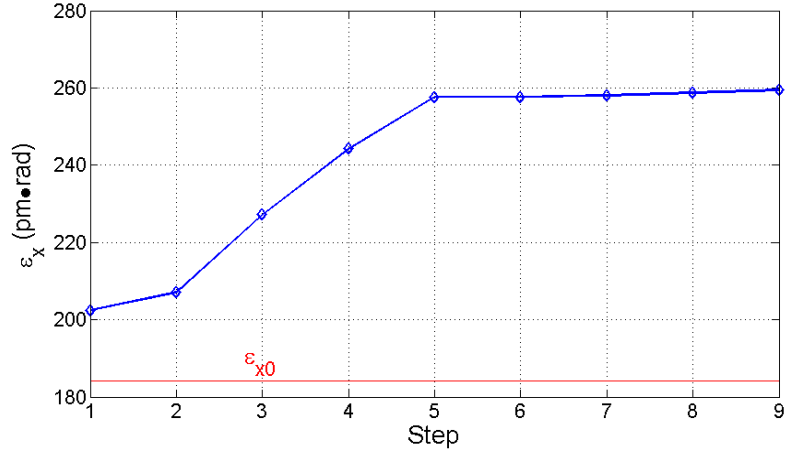

Fig 7. Equilibrium horizontal emittance fluctuates during the injection process caused by the IBS effect.

During the injection process, not only the bunch length is changed but also the RF acceptance for the stored beam is modified. The reduced RF acceptance will affect the beam lifetime. Fig. 8 shows the evolution of the energy acceptance of the main RF bucket. At step 2 in Fig. 8, a sudden drop of the energy acceptance for the main RF bucket is found at which time the second RF bucket is created. At this step, the particles in the stored beam with energy deviation larger than the main RF acceptance are not lost directly because the total RF acceptance is still large as shown in Fig. 9. Those particles escaped from the main RF buckets will be captured by the created RF buckets or experience a large oscillation in the total RF buckets.

The minimum energy acceptance in Fig. 8 is 0.44%, which is consistent with tracking results shown in Fig. 4. The reduced main RF bucket energy acceptance will reduce the quantum and the Touschek beam lifetime of the stored beam. The quantum lifetime is about 0.5 hours. The Touschek lifetime evaluated by the ZAP [23] is 2 minutes. As explained above, the lost particles will be caught by the created buckets or experience a large oscillation in the total RF bucket, so the instantaneous beam lifetime drop will not be observed. Assuming step 1 to step 2 takes 1 second, only 0.8% of total charge escapes the main RF buckets. When on axis injection performs, they will be swapped-out appears like an injection efficiency decrease. The minimum energy acceptance of the main RF bucket is relevant to the synchrotron phase deviation $\Delta\theta$. The relationship is shown in Fig. 10. A reasonable $\Delta\theta$ is a compromise of the injection efficiency and the energy oscillation during dumping process. For the case of SSRF-U, 12-degree is the optimized value of the $\Delta\theta$ to achieve a good injection efficiency while get small energy oscillation of the injection beam as shown in Fig. 10.

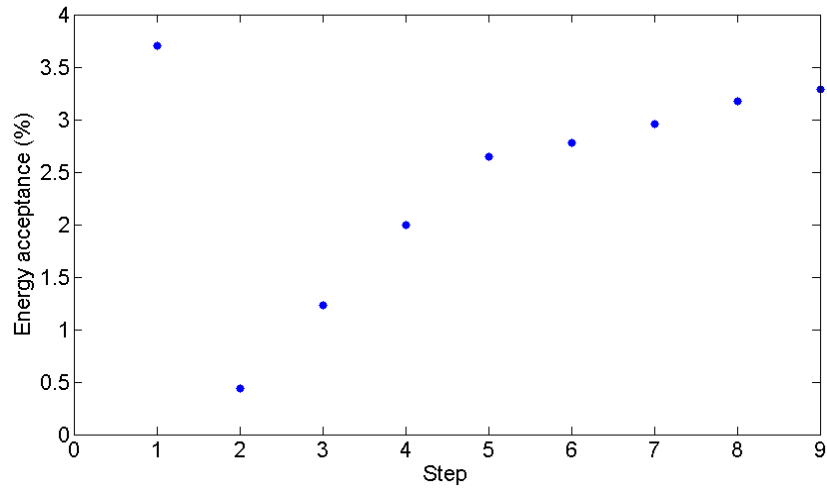

Fig 8. Energy acceptance evolution.

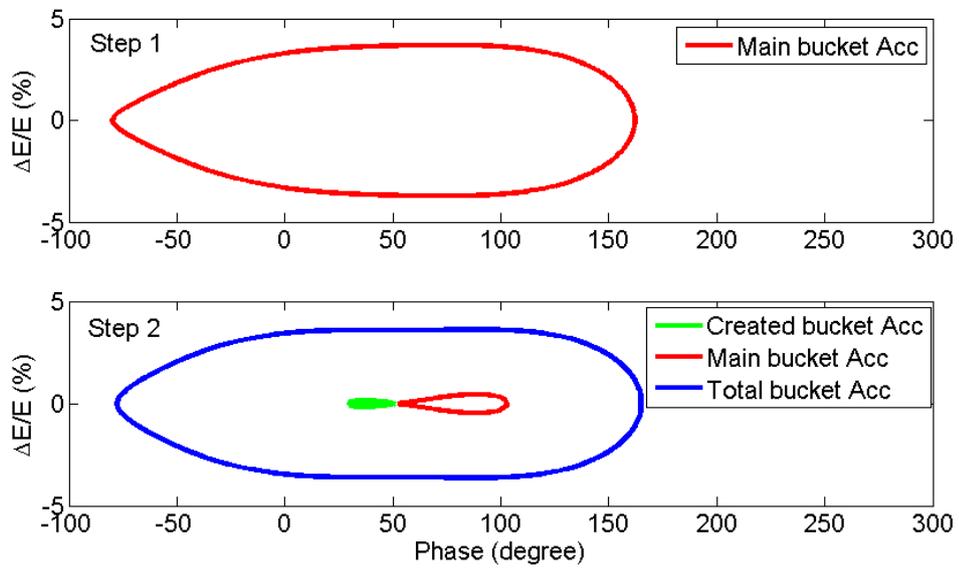

Fig. 9 RF acceptance evolution at the step of the second RF bucket is created.

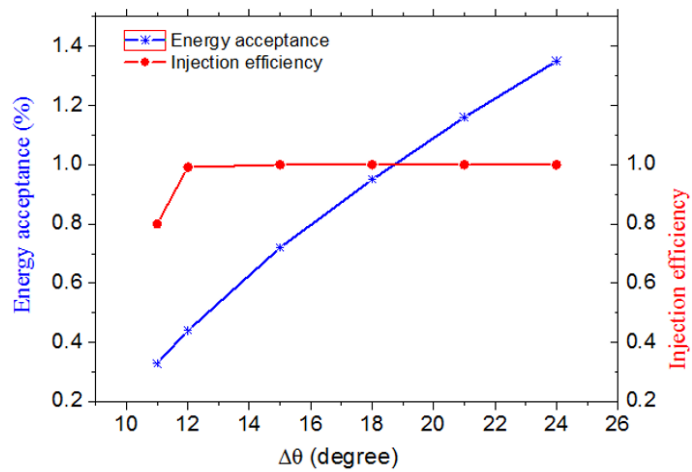

Fig. 10. Main RF bucket energy acceptance and injection efficiency Vs. phase deviation for SSRF-U

## V. DISCUSSION

This injection scheme relies on alteration two RF system voltages. A big amplitude RF voltage modulation needs a large detuning angle change. This is the reason we propose an alteration period in the seconds range, leaving enough time for the detuning process. And zero voltage of both RF system should be avoided.

Actually this scheme will become very complicated when a 3rd harmonic cavity system is employed to lengthen the bunch which is usually considered to decrease the IBS effect as well as to improve the Touschek beam lifetime in low emittance storage ring. For this purpose, it can be found out in Fig. 6 that the 2nd harmonic cavity can act as a bunch lengthening system. On the other hand using an active 3rd harmonic cavity to build up a double-frequency RF system for on axis injection is possible with the cost of an even more challenging fast kicker.

It is mentionable that gap changes of insertion devices (IDs) will greatly affect low emittance storage ring performances as pointed out in [21]. The interplay of the ID gap change and the RF synchrotron phase shift will also affect this injection scheme. As suggested in [21] using a damping wiggler to compensate the radiation loss change caused by ID gap adjustment should be considered to stable the synchrotron phase.

For the initial injection, a cycle of injection takes minutes, and for the top-up injection the refilling period is also in minutes, so a few seconds extra RF voltage alteration period for the injection will not significantly lengthen the overall injection time.

Beam transient loading is another concern for keeping a constant synchrotron phase deviation $\Delta\theta$ between the head and the tail bunches in a bunch train [24]. Transient beam loading is evaluated by the following formula [25].

$$\Delta\varphi \approx \frac{I}{2V_c}\frac{R_s}{Q}\omega_{rf}\Delta t, \quad (6)$$

Where $I$ is the beam current, $\Delta\varphi$ is the phase shift between the head and tail bunches, $V_c$ is the RF voltage, $R_s$ is the shunt impedance of the cavity, $Q$ is quality factor, and $\Delta t$ is the time of the empty gap. The formula is valid when $T_0 \ll T_f$, $T_0$ is the revolutionary period, $T_f = \frac{2Q}{\omega_c}$. With SSRF superconducting cavity parameters $Q=1.7x10^5$, $R_s/Q=90$. $\Delta\varphi \approx \frac{I}{2V_c}\frac{R_s}{Q}\omega_{rf}\Delta t \approx 0.004 = 0.23°$ which can be neglected.

## ACKNOWLEDGMENT

The authors would like to thank Dr. Chunguang Jing and Dr. Paul Schoessow for help composing the manuscript. This work was supported by the Youth Innovation Promotion Association of Chinese Academy of Sciences No. 2011196.

## REFERENCE

[1] Einfeld D, Schaper J and Plesko M, in *Proceedings of the Particle Accelerator Conference, PAC95*,

Dallas, TX (IEEE, New York, 1995), pp 177–179.

[2] Leemann S C, Andersson A, Eriksson M, Lindgren L J, Wallén E, Bengtsson J and Streun A, Phys. Rev. ST Accel. Beams **12**, 120701 (2009).

[3] M. Borland, in *11th International Conference on Synchrotron Radiation Instrumentation (SRI 2012)*, IOP Publishing, Journal of Physics: Conference Series **425** (2013) 042016.

[4] M. Bei, M.Borland, Y.Cai, P.Elleaume, R.Gerig, K.Harkay, L.Emery, A.Hutton, R.Hettel, R. Nagaoka, D.Robin, C.Steier, Nucl. Instrum. Methods Phys. Res., Sect. A **622** (2010) 518–535.

[5] Yunhai Cai, Karl Bane, Robert Hettel, Yuri Nosochkov, Min-Huey Wang, Phys. Rev. ST Accel. Beams, **15**, 054002 (2012).

[6] D. Einfeld, M. Plesko, Proc. SPIE 2013, San Deigo, CA, USA, (1993) 201-212.

[7] D. Kaltchev, R.V. Servranckx, M.K. Crddock, and W. Joho, in Proceedings of the Particle Accelerator Conference 95, Dallas, TX, U.S.A. (1995) 2823-2825.

[8] H. Tarawneh, C. Steier, R. Falcone, D. Robin, H. Nishimura, C. Sun, W. Wan, in *17th Pan-American Synchrotron Radiation Instrumentation Conference (SRI2013)*, IOP Publishing, Journal of Physics: Conference Series **493** (2014) 012020.

[7] Yipeng Sun, Michael Borland, in *Proceedings of the 25th Particle Accelerator Conference, PAC-2013*, Pasadena, CA, USA, (IEEE, New York, 2013), 258-260.

[10] Y. Shimosaki, K. Soutome, J. Schimizu, K. Kaneki, M. Takao, T. Nakamura, H. Ohkuma, in *Proceedings of the 2nd International Particle Accelerator Conference, IPAC2011*, San Sebastián, Spain, (EPS-AG, Spain, 2011), 942-944.

[11] M. Borland, Nucl. Instrum. Methods Phys. Res., Sect. A **557** (2006), 230-235.

[12] K. Tsumaki, N. Kumagai, Nucl. Instrum. Methods Phys. Res., Sect. A **565**, (2006) 394-405.

[13] R. Abela, et al., *Third European Particle Accelerator Conference, EPAC92*, Berlin, Germany, (1992) 486-488.

[14] L. Emery, M. Borland, *in Proceedings of the 2003 Particle Accelerator Conference (2003)*, Portland, OR, U.S.A., 256-258.

[15] A. Xiao, M. Borland, C. Yao, in *Proceedings of the 25th Particle Accelerator Conference, PAC-2013*, Pasadena, CA, USA, (IEEE, New York, 2013), 1076-1078.

[16] *M. Aiba, M. Böge, F. Marcellini, Á. Saá Hernández, and A. Streun, Phys. Rev. ST Accel. Beams, 18, 020701 (2015).*

[17] G. N. Kulipanov, in *Proceedings of the 8th International Conference on High-Energy Accelerators,* Geneva, Switzerland, 1971, 138-139.

[18] T. Nakamura, in *Proceedings of the 2nd International Particle Accelerator Conference, IPAC2011*, San Sebastián, Spain, (EPS-AG, Spain, 2011),1230-1232

[19] S.Q. Tian, M.Z. Zhang, Q.T. Zhang, B.C. Jiang, Z.T. Zhao, in Proceedings of IPAC2015, Richmond, VA, U.S.A. (2015), 304-306.

[20] Z.T. Zhao, L.X. Yin, Y.B. Leng, B.C. Jiang, S.Q. Tian, M.Z. Zhang, in Proceedings of IPAC2015, Richmond, VA, U.S.A. (2015), 1672-1674.

[21] S.C. Leeman, Phys. Rev. ST Accel. Beams, **17**, 050705 (2014).

[22] Hans Grote, et al., The MAD-X Program User's Reference Manual, ttp://madx.web.cern.ch/madx/

[23] Michael S. Zisman, Swapan Chattopadphyay and Joseph J. Bisognano, LBL-21270 (1986).

[24] J. M. Byrd, S. De Santis, J. Jacob, and V. Serriere, Phys. Rev. ST Accel. Beams, **5**, 092001 (2002).

[25] D. Boussard, in *Proceedings of the 1991 Particle Accelerator Conference, PAC91,* San Francisco, (1991) p2447.